# Highlights

**Mental health of computing professionals and students: A systematic literature review**

Alicia Julia Wilson Takaoka, ,Kshitij Sharma

- We identified three different themes from the state-of-the-art: General Exploration, Design and Develop, and Test and Improve.

- One of the key challenges is that many studies have reported low sample sizes, which poses challenges to reproducibility and generalizability of results as well as threats to construct validity.

- The current state-of-the-art has few intervention-based studies, especially targeted for cs students and professionals; however, there is a vast opportunity for the development of interventions targeting specifically the mental health of professionals and students from the field of computing.

# Mental health of computing professionals and students: A systematic literature review[*],[**]

Alicia Julia Wilson Takaoka, [a],[*],[1], Kshitij **Sharma**[b]

[a]*Norwegian University of Science and Technology, Trondheim, 7034, Norway*
[a]*Norwegian University of Science and Technology, Trondheim, 7034, Norway*

**ARTICLE INFO**

**ABSTRACT**

The intersections of mental health and computing education is under-examined. In this systematic literature review, we evaluate the state-of-the-art of research in mental health and well-being interventions, assessments, and concerns like anxiety and depression in computer science and computing education. The studies evaluated occurred across the computing education pipeline from introductory to PhD courses and found some commonalities contributing to high reporting of anxiety and depression in those studied. In addition, interventions that were designed to address mental health topics often revolved around self-guidance. Based on our review of the literature, we recommend increasing sample sizes and focusing on the design and development of tools and interventions specifically designed for computing professionals and students.

## 1. Introduction

Mental health is a growing area of concern for computer science in academia and industry. In part, this was caused by isolation and compounding concerns and stressors due to the COVID-19 pandemic (Williams, Pinto, Abrahim, Bhattacharya, Harrison, Burton, McIntyre, Negi, and Hayes (2023)), but the need for mental health and well-being support was recognized and documented before the pandemic. Some causes of mental health distress include globalization Colantone, Crino, and Ogliari (2019), time constraints in the workplace (Carroll and Conboy (2020); Conboy, Dennehy, and O'Connor (2020)), toxic environments (Singh, Bongiovanni, and Brandon (2021)), and demanding workloads (Pimenta, Gonçalves, Carneiro, Fde-Riverola, Neves, and Novais (2015)).

Meaningful research about the mental health and well-being has occurred either outside the discipline of computer science or non-student populations. In psychiatry and psychology, work has been done to identify connections between stigma and disclosure and develop digital self-mitigated interventions. Ho, Potash, Fong, Ho, Chen, Lau, Yeung, and Ho Ho, Potash, Fong, Ho, Chen, Lau, Yeung, and Ho (2015) addressed the complex relationship between stigma, self-esteem, and depression in students in Hong Kong. They point out that complexities around disgracing the family name and stereotypes about people with mental illness cause many to not seek support and self-stigmatize. They examined different scales for evaluating stigmas and found strong associations between feeling discrimination and the decision to not disclose mental illness to others. In Australia, researchers worked to design apps to mitigate symptoms of depression when other support is unavailable.

Bakker, Kazantzis, Rickwood, and Rickard Bakker, Kazantzis, Rickwood, Rickard et al. (2016) reviewed the landscape of apps for identifying and mitigating changes in mental health. They identified sixteen recommendations for designing mental health applications for identifying and designing applications using therapeutic conventions, which they employed in their own application development. Through monitorng engagement Bakker and Rickard (2018) and

---

[*]This document is the results of the research project funded by the National Science Foundation.

[**]The second title footnote which is a longer text matter to fill through the whole text width and overflow into another line in the footnotes area of the first page.

    This note has no numbers. In this work we demonstrate $a_b$ the formation $Y\_1$ of a new type of polariton on the interface between a cuprous oxide slab and a polystyrene micro-sphere placed on the slab.

[*]Corresponding author

[**]Principal corresponding author

  ✉ alicia.j.w.takaoka@ntnu.no (A.J.W.T. ); kshitij.sharma@ntnu.no (K. Sharma)

ORCID(s): 0000-0001-9014-3301 (A.J.W.T. )

[1]This is the first author footnote. but is common to third author as well.

[2]Another author footnote, this is a very long footnote and it should be a really long footnote. But this footnote is not yet sufficiently long enough to make two lines of footnote text.

---

comparative analysis of three applications Bakker, Kazantzis, Rickwood, and Rickard (2018), the researchers were able to develop a tool that uses cognitive behavior change to predict changes in mental health of regular users of their app Bakker and Rickard (2019). This tool was also useful in the design of a preliminary mental health literacy assessment of app users.

Since the COVID-19 pandemic, examining mental health, well-being, workloads, and needs to foster well-being in computing education and the software development workplace has exploded. Many researchers have realized that the toxicity endured in some software engineering workspaces Singh et al. (2021) and the needs developers have Yacoby, Girash, and Parkes (2023); Williams et al. (2023) must be met in order to foster positive team dynamics. Some researchers and others in the dev community recognize that being a safe space for developers with neurodiversity and being neuroinclusive are integral to a successful working environment Farsinejad, Russell, and Butler (2022). This work is complimented by a pair of studies that use existing technologies to track and predict changes in mental health and well-being.

Kumar, Sharma, and Sharma (2021) identified using sensors in the Internet of Things model to help individuals recognize changes in their mental well-being through wearable technology. Using biobehavior markers like pulse, body temperature, and breathing, the reearchers designed a model for tracking and predicting changes using a neural network. Another study using the Internet of Things employs chatbots for student interventions.

Kumar, Yu, Chung, Shi, and Williams (2022) investigated chatbots as a possible mental health intervention in working with students. In their study, students interacted with three types of large language model chatbots. They found that the friend chatbot and the coach chatbot mimicked expected behaviors of their roles and may indicate a need for further evaluation. However, these students were not solely in the discipline of computer science or computing-adjacent fields.

In the scope of mental health and well-being literature about computing education, some previous literature reviews have addressed the needs of students and the supports professors can offer Akullian, Blank, Bricker, DuHadway, and Murphy (2020) and how to identify mental health needs using social media cues Chancellor and De Choudhury (2020) have been published, but none have taken a holistic view of examining the state of the mental health of computer science students in university, assessments used in analysis, and technologies and tools designed to address the needs of computer science students.

Still, the landscape of mental health and well-being in university computer science students is under-studied. In this systematic literature review, we seek to address the following research questions:

**RQ1:** What is the current status of mental health and university-level computer education research?

- **RQ1.1:** Which mental health assessments are used in university-level computer education research?

- **RQ1.2:** What technologies or interventions have been developed to support university-level computer education?

- **RQ1.3:** What types of data do these assessments, interventions, and technologies capture?

- **RQ1.4:** What insights come from mental health technologies and interventions for university-level computing education?

- **RQ1.5:** What are the gender differences present in empirical results in mental health and university level computing education research?

**RQ2:** What are the challenges, opportunities, and limitations of mental health technologies and interventions in computing education research?

In addressing these questions, we will identify hypotheses for future evaluation and with computer science university students as the target population and provide an overview of the tools that have been developed to address the needs of these students in a post-pandemic university setting.

## 2. Methods

To minimize potential biases (researchers) and support reproducibility (especially in the areas of software engineering and information systems) in this systematic review, we follow a transparent and widely accepted process. In addition to minimizing bias and supporting reproducibility, systematic reviews provide information about the impact of a phenomenon across a wide range of settings, contexts, and empirical approaches. As a result, systematic reviews can provide evidence that the phenomenon is robust and transferable if the selected studies give consistent results (Kitchenham & Charters, 2007).

## 2.1. Article Collection

Several procedures were followed to ensure a high-quality literature review on mental well-being and computing education. A comprehensive search of peer-reviewed articles was conducted in June 2023 (short papers, posters, dissertations, editorials and reports were excluded). The following search phrase was used.

*HCI OR AI OR Informatics + "mental health"*

Publications were selected from 2000 onward because there have been tremendous technological and societal advances since 2000 for supporting issues related to mental well-being. Various databases were searched, including SpringerLink, Wiley, Assembly of Computer Machineries [ACM] Digital Library, IEEE Xplore, Science Direct, SAGE and ERIC. The search process uncovered 4048 peer-reviewed articles.

## 2.2. Inclusion and Exclusion Criteria

The selection phase determines the literature review's overall validity, and thus, it is essential to define specific inclusion and exclusion criteria. We applied eight quality criteria informed by related works (e.g., Dybå & Dingsøyr, 2008), followed by two filters to include/exclude the articles from this review. The selection phase determines the overall validity of the literature review, and thus it is important to define specific inclusion and exclusion criteria. As Dybå and Dingsøyr (2008) specified, the quality criteria needs to cover three main issues (i.e., rigour, credibility, and relevance) that need to be considered when evaluating the quality of the selected studies. We applied eight quality criteria informed by related works (e.g., Dybå & Dingsøyr, 2008). The following are the criteria:

1. Does the study clearly address the research problem?
2. Is there a clear statement of the aims of the research?
3. Is there an adequate description of the context in which the research was carried out?
4. Was the research design appropriate to address the aims of the research?
5. Does the study clearly determine the research methods (subjects, instruments, data collection, data analysis)?
6. Was the data analysis sufficiently rigorous?
7. Is there a clear statement of findings?
8. Is the study of value for research or practice?

The first filter consisted of the following rules:

1. Papers should be in English only.
2. Papers should be published in peer-review venues only (i.e., book chapters, opinion articles, editorials, and magazine articles were removed).
3. Papers should have more than 5000 words, because short papers lack a substantial contribution to the field.
4. Papers should not be duplicated.

After applying the first, there were 635 papers left, which were subjected to the second filter. The second filter was based on reading the title and the abstracts of the papers and had the following rules:

1. Papers should present empirical data.
2. Review papers should be removed.
3. If the paper does not include aspects related to MWB or computer science education or computing education, it should be removed.
4. If the paper presents a system that has no potential aspect of addressing the MWB of computing students, it should also be removed.

After applying the second filter, there were 39 papers left. These papers were then subjected to the data analysis. While analysing the papers, we observed a further 10 papers that did not fulfill the inclusion criteria. Therefore, the results presented in this systematic literature review are from the remaining 28 papers. Upon reviewing these papers, we identified one paper that was published in two different venues. It was subsequently omitted.

Figure 1 shows the systematic filtering of publications. In this table, we identify the databases and the total papers returned from each query. We also identify the results after each round of filtering.

| Databases | Preliminary | Round 1 | Round 2 |
|---|---:|---:|---:|
| ACM | 40 | 21 | 11 |
| Science Direct | 631 | 500 | 37 |
| IEEE | 48 | 22 | 42 |
| Wiley | 0 | 0 | 0 |
| SAGE | 0 | 0 | 0 |
| Web of Science | 19 | 16 | 16 |
| Springer | 3310 | 76 | 23 |
| Snowball | 0 | 0 | 44 |
| Totals | 4048 | 635 | 173 |

**Figure 1:** Systematic Filtering of Publications

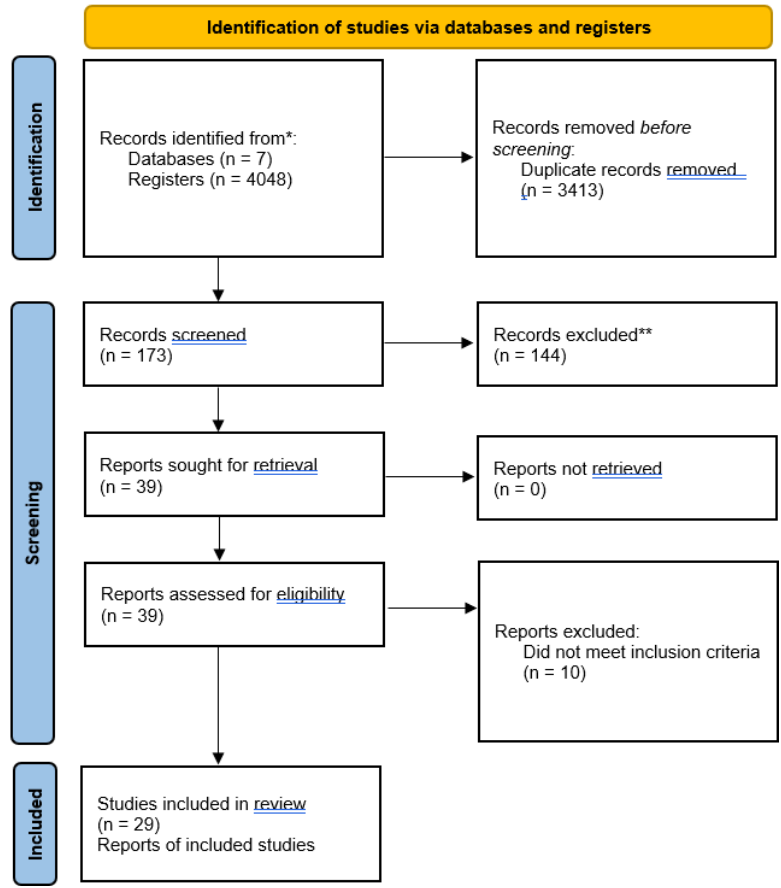

**Figure 2:** Systematic Filtering of Publications

Figure 2 highlights the filtering process in an aggregated state as recommended using PRISMA[3] guidelines for systematic and meta-analyses. In addition to the initial filtering rounds, the figure highlights the screening process and the total number of included reports.

[3] http://prisma-statement.org/prismastatement/flowdiagram.aspx?AspxAutoDetectCookieSupport=1

## 2.3. Data Analysis

In total, 28 studies were found to meet the quality criteria, which is as we have stated above. These studies have been coded according to specific areas of focus in which they have been conducted. It was through this process that we were able to consolidate the essence of the studies as well as the main focus of them. It was decided that the categories selected should represent the MWB and technological/societal aspects of the paper as well as its objectives and content. By categorizing the papers in our literature review, we were able to record all the necessary information from the papers in our literature review and use it to address the research questions we had in mind. In particular, each of the collected studies was analyzed according to the following elements:

1. number of participants
2. gender breakdown
3. educational major
4. country
5. continent
6. specific course
7. who was the intervention for
8. how information was provided or technology involved (website, app, )
9. what was the intervention
10. what was the intervention for (if there was a specific mental health issue)
11. research method (qual/quant/mixed)
12. research method (survey, interview, …)
13. Analysis
14. measurements/dependent variable
15. theories used
16. results
17. if there is any gender specific implication
18. research questions and objectives
19. implications

In the next section, we present the results of the analysis using common bibliomentric techniques and thematic analysis. We provide an overview of the studies and the context. We then address the interventions followed by measurements and analysis. We then introduce the mental health aspects of the papers, technologies and interventions developed to address mental health topics. Then, the general implications and gender-based implications are presented. Finally, we address the opportunities and challenges identified from this systematic review.

## 3. Results

We identified and examined 28 articles for this systematic literature review based on our inclusion and exclusion criteria. Those papers are:

1. An Analysis of the Psychological Implications of COVID-19 Pandemic on Undergraduate Students and Efforts on Mitigation Rao, Pushpalatha, Sapna, and Monika Rani (2021)
2. Assessing Mental Stress Based on Smartphone Sensing Data: An Empirical Study Wang, Wang, Wang, Xiong, Zhao, and Zhang (2019)
3. Association of Positive and Negative Feelings with Anxiety and Depression Symptoms among Computer Science Students during the COVID-19 Pandemic Passos, Murphy, Chen, Santana, and Passos (2022)
4. Covariation of Depressive Mood and Spontaneous Physical Activity in Major Depressive Disorder: Toward Continuous Monitoring of Depressive Mood Kim, Nakamura, Kikuchi, Yoshiuchi, Sasaki, and Yamamoto (2015)
5. Depression and anxiety among online learning students during the COVID-19 pandemic: a cross-sectional survey in Rio de Janeiro, Brazil Pelucio, Simões, Dourado, Quagliato, and Nardi (2022)
6. Empowering First-Year Computer Science Ph.D. Students to Create a Culture that Values Community and Mental Health Yacoby et al. (2023)

7. Engagement with a cognitive behavioural therapy mobile phone app predicts changes in mental health and wellbeing: MoodMission Bakker and Rickard (2019)

8. Engagement in mobile phone app for self-monitoring of emotional wellbeing predicts changes in mental health: MoodPrism Bakker and Rickard (2018)

9. Hierarchical deep neural network for mental stress state detection using IoT based biomarkers Kumar et al. (2021)

10. Identifying the Prevalence of the Impostor Phenomenon Among Computer Science Students Rosenstein, Raghu, and Porter (2020)

11. The Impact of COVID-19 on the CS Student Learning Experience: How the Pandemic Has Shaped the Educational Landscape Siegel, Zarb, Anderson, Crane, Gao, Latulipe, Lovellette, McNeill, and Meharg (2022)

12. The Impact of virtual learning on students' educational behavior and pervasiveness of depression among university students due to the COVID-19 pandemic Azmi, Khan, and Azmi (2022)

13. Investigating the impact of the COVID-19 pandemic on computing students' sense of belonging Mooney and Becker (2021)

14. Is screen time associated with anxiety or depression in young people? Results from a UK birth cohort Khouja, Munafò, Tilling, Wiles, Joinson, Etchells, John, Hayes, Gage, and Cornish (2019)

15. Negative influences of Facebook use through the lens of network analysis Faelens, Hoorelbeke, Fried, De Raedt, and Koster (2019)

16. Predicting Melancholy risk among IT professionals using Modified Deep Learning Neural Network (MDLNN) Rosaline, Nasreen, Suganthi, Manimegalai, and Ramkumar (2022)

17. The Prevalence of Anxiety and Depression Symptoms among Brazilian Computer Science Students Soares Passos, Murphy, Zhen Chen, Gonçalves de Santana, and Soares Passos (2020)

18. Prevalence and Psychological Effects of Hateful Speech in Online College Communities Saha, Chandrasekharan, and De Choudhury (2019)

19. A randomized controlled trial of three smartphone apps for enhancing public mental health Bakker et al. (2018)

20. Role Modeling as a Computing Educator in Higher Education: A Focus on Care, Emotions and Professional Competencies Grande, Kinnunen, Peters, Barr, Cajander, Daniels, Lewis, Sabin, Sánchez-Peña, and Thota (2022)

21. Self-esteem and socialisation in social networks as determinants in adolescents' eating disorders Frieiro, González-Rodríguez, and Domínguez-Alonso (2022)

22. 'Smartphine': Supporting students' well-being according to their calendar and mood Baras, Soares, Paulo, and Barros (2016)

23. A Study of The Effects of Short-Term AI Coding Course with Gamification Elements on Students' Cognitive Mental Health Kamarudin, Ikram, Azman, Ahmad, and Zainuddin (2022)

24. Sustaining Student Engagement and Equity in Computing Departments During the COVID-19 Pandemic Thiry and Hug (2021)

25. Take a deep breath: Benefits of neuroplasticity practices for software developers and computer workers in a family of experiments Penzenstadler, Torkar, and Martinez Montes (2022)

26. Taking the pulse of nations: A biometric measure of well-being Blanchflower and Bryson (2022)

27. User's perception on mental health applications: a qualitative analysis of user reviews Thach (2018)

28. Virtual Reality Course Based on the SAT Counseling Method for Self-Guided Mental Healthcare Ito, Kamita, Matsumoto, Munakata, and Inoue (2018)

Of those, 35 study sites were identified. In addition, one study defined the study site as International, two did not explicitly mention a study site, and three described the study site as online. A may of the study sites by country can be seen in Figure 3.

The context of these studies spans from student, faculty, and administrative interventions to developing tools and using assessments to identify the current mental health status of students in computer science education. In the next section, we will provide a deeper look into the context of these studies.

## 3.1. Context of the studies

This section presents the state-of-the-art from the research context and design perspectives. This includes the number of participants, their gender breakdown, education major of the participants, country, and the specific CS course. We observe that the maximum number of studies (14) used a sample between 100 and 500 participants. Further, there were five studies who used the sample between 500 and 1000 participants, four studies used less than 25 participants, two studies used between 25 and 50 participants. There were two studies that used more than 1000 participants and another two studies that used more than 5000 participants. One study did not mention the number of participants included in the contribution. Concerning the gender ratio between females and males four studies reported between 70% and 80% of the participants to be females, three studies reported less than 10% of females while other three studies reported the female percentage should be between 20% and 30%. Furthermore there were two studies that reported female percentage to be between 30% and 40% other two studies reported the female percentage between 40% and 50% and another two studies reported female percentage to be between 50% and 60%. The female participation was between 10% and 20%, 80% and 90%, 90% and 100% for one study each. There were ten studies that did not report this percentage and there was only one study that also considered a wider gender spectrum. In terms of reporting the student population oh majority of studies (12) included university participants without specifying further details about their educational levels. There were two studies which specifically included undergraduate students, while primary and secondary students were used by one study each. There was only one study that used computer science teachers as their participants. Finally there were 11 cities that did not report the educational level of their participants. Based on the geographical distribution of the studies six studies were situated in USA while other four studies reported international participants, and two sets of four studies were situated in UK and Brazil, respectively. Apart from this, we observed the following countries included in three contributions each – Australia and Japan. There were two studies included participants from India. Finally, the following countries included in one contribution each Canada, Ireland, Malaysia, Pakistan, Saudi Arabia, Scotland, Spain, Switzerland. There were two studies that did not report the nationality of their participant.

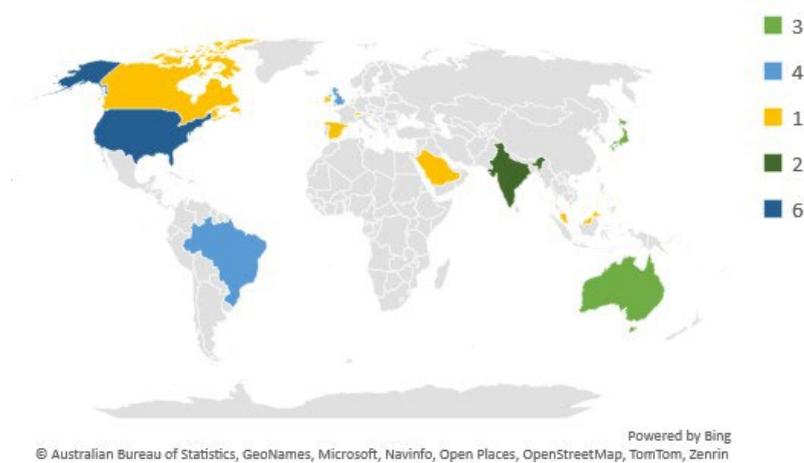

**Figure 3:** Study locations

## 3.2. Interventions used in the studies

This section presents the state-of-the-art from the intervention perspectives. This includes the target audience of the interventions, how the interventions were delivered, and what modes were used to deliver the interventions. A majority of studies (16 out of 28) used students from computer science departments has the target audience of their interventions. Twisted is used both students and non students and two studies used professionals as target audience. The following category of target audience was used by one study each doctoral students, matriculated students, young students, and educators. There were four studies that did not explicitly mentioned the exact audience of their interventions. Considering the medium of communication with audience nine studies used applications, while seven studies used emails. There were four studies that use social media platform to communicate with their target audience while three studies used physical interventions and three other studies used web apps for communication purposes. The following communication channels were used by 1 study each chatbots and virtual reality applications. Six studies that did not explicitly mention the communication medium used for interventions. Finally considering the delivery modes for intervention continuous monitoring and surveys were used by seven studies each three studies used online learning while following delivery modes were used by one study each, cognitive behavioral therapy, forums, games, physical exercise, situated awareness tests, and self regulated learning. Six studies that did not explicitly mention the delivery modes used for interventions. Figures 4 and 5 show the collection of instruments used in the studies referenced

| Emotional Assessments (3) | Observational Assessments (7) | Technological Assessments (5) |
|---|---|---|
| Care score | continuous monitoring | app engagement scale |
| Emotion score | demographic information | gamified coding exercises |
| Emotional Self-Awareness Scale – Revised | logging the "mood" | general mental health app based intervention |
| | role modeling using scaffolded reflection framework | multilingual chatbot |
| | self-reporting | Online learning |
| | specific breathing practice -- related to neuroplasty | |
| | working with tutor | |

**Figure 4:** Emotional, Observational, and Technological Assessments found in the studies

| Psychological Assessments (30) | |
|---|---|
| 14 category psychological scale | OSMI mental health survey |
| 30 temperament measures | Patient Health Quality Questionnaire (PHQ-9) |
| Beck Anxiety Inventory | Perceived Productivity instrument (HPQ) |
| Beck Depression Inventory | Positive Thinking Scale (PTS) |
| Clance Imposter Syndrome Scale | Psychological Well-Being scale (PWB) |
| Cognitive Behavior Thory | revised Clinical Interview Schedule (CIS-R) |
| Coping Self-Efficacy Scale | Rosenberg Self-Esteem Scale |
| DASS-21 e-questionnaire set | Scale of Positive and Negative Experience (SPANE) |
| Eating Attitudes Test-26 (EAT-26) | Self-Efficacy instrument |
| ESOC-39 scale | Stress Scale 21 |
| Generalised Anxiety Disorder Scale 7-Item (gad-7) | Structured Association Technique |
| Hamilton depression rating scale | Trier Social Stress Test |
| Hospital Anxiety and Depression Scale | Warwick-Edinburgh Mental Well-being Scale |
| Mental Health Literacy Questionnaire | Well-Being Assessment (DAWBA) |
| Mindfulness Attention Awareness Scale (MAAS) | Zung (Self-Rating Depression Scale) questionnaire |

**Figure 5:** Psychological Instruments

## 3.3. Measurements and Analysis

This section presents the state-of-the-art from the measurements collected and analysis conducted in the studies. This includes the analysis method, mode of data collection, the type of statistical analysis Conducted, and the dependent variables used in the studies. We observed that 19 out of 28 studies used quantitative research methods while nine had a mixed method approach to their research design and there was only one study that used a qualitative research approach. When it comes to data collection modes a huge majority, that is 25 out of 28 studies, used as survey while three other studies used interviews. The following data collection modes were used by two studies each biomarkers, diary, sensor data, text data; while the following mouths were used by one study each assessment, technology evaluation, and data logs. Considering the statistical method used in the analysis, we observed that 14 studies used inferential statistics while ten studies used descriptive analysis. On the other hand eight studies used regressions and six studies used predictive analysis. Content analysis, grounded theory approach, thematic analysis were used by three studies each. Finally natural language processing social network analysis and visualizations were used as a statistical analysis method in one study each. From the point of view of main dependent variable of the studies depression and anxiety combined and overall mental health were used as a dependent variable in 8 studies each, stress was used as the main dependent variable in seven studies. There were five studies that used depression as the main dependent variable while there were four studies that used anxiety as their main dependent variable. Imposter syndrome and self esteem were the primary dependent variables of two studies each and the following aspects of mental health were used as a primary dependent variable in

| Mental Health Topic | Count |
|---|---|
| Anxiety | 13 |
| Attention Awareness | 1 |
| Care | 1 |
| Community | 1 |
| Depression | 13 |
| Eating Disorders | 1 |
| General Mental Well-Being | 3 |
| Imposter Syndrome | 2 |
| Isolation | 1 |
| Melancholy | 1 |
| Overall Mental Health | 2 |
| Productivity | 2 |
| Role Modeling | 1 |
| Self-Esteem | 2 |
| Self-Efficacy | 1 |
| Self-Guided Mental Healthcare | 1 |
| Sense of Belonging | 1 |
| Sleep | 1 |
| Stress | 7 |
| Stressors | 1 |
| Well-Being | 2 |

**Table 1**
Mental health topic and total number of papers that address it, in alphabetical order one study each: care and support, eating disorders, intersectionality, isolation, melancholy, mood, role modeling, self efficacy, and sense of belonging.

### 3.4. Mental health aspects of the papers

In this systematic literature review, 21 mental health topics were addressed. Anxiety and depression were both present in 13 papers while stress was addressed in 7 papers, and general well-being was addressed in 3 papers. The other topics were present in either one or two studies. A full list can be seen in Figure 1.

The theories used to construct the theoretical framework in which the studies were situated were evaluated. A total of 23 theories were found across the papers. These can be seen in Figure 6. Surprisingly, fourteen papers omitted mention of a guiding theory or theoretical framework. This was the overwhelming majority, representing 50% of our sample. With the exception of Cognitive Behavioral Theory, or CBT, which was present in three studies, each remaining study constructed a theoretical framework that was unique to its needs. Therefore, there is no standardized theory used to address or evaluate mental health in computing education.

Prior to this exploration, we were expecting to see some theories emerge with consistency in this study. Based on our experiences with computing education and informatics research, we were expecting CBT, which was present. We were also expecting gamification and other theories to lead the theoretical choices in this systematic review. A full list of our expected theories that were not present in these studies is shown in Figure 7.

After analyzing the theories used in the sample, we assessed the technologies and interventions developed to address mental health topics in computer science education.

### 3.4.1. Technologies and Interventions

Several papers in our study formed a subset of designed technology-based applications (app) and adaptations to address mental health topics. Bakker and Rickard developed a series of mental health applications in relation to their work. MoodMission , MoodPrism, and MoodKit Bakker and Rickard (2019), Bakker and Rickard (2018), Bakker et al. (2016), Bakker et al. (2018) were designed to address different types of intervention needs in tandem or based on specific goals. MoodKit[4] is a journaling tool and tracker designed to provide insights into understanding and changing the user's mood using CBT. MoodMission[5] is created using evidence-based research as a self-guided intervention when the user experiences depression and anxiety. The third app in this series is MoodPrism [6]. MoodPrism is designed to help increase the user's emotional awareness and well-being as a self-guided intervention. These three products are on the market for Android and iPhone today.

---

[4]https://www.thriveport.com/products/moodkit/
[5]https://moodmission.com/

| Theoretical framework | Papers |
|---|---|
| None explicitly stated | 14 |
| care theory | 1 |
| cbt | 3 |
| cohort-building design | 1 |
| contingent self esteem | 1 |
| ecological momentary assessment | 1 |
| emotion theory | 1 |
| equity-oriented engagement | 1 |
| global self esteem | 1 |
| grounded theory | 1 |
| high impact practices | 1 |
| imposter phenemenon | 1 |
| inductive reasoning | 1 |
| intersectionality | 1 |
| mindfulness | 1 |
| mindfulness attention awareness | 1 |
| professional competencies framework | 1 |
| role modeling framework | 1 |
| self regulatory competance | 1 |
| self-efficacy | 1 |
| sense of belonging | 1 |
| space effect | 1 |
| structured association tech | 1 |
| theory of stressors | 1 |

**Figure 6:** Theories in use

| Theories we expected but were not present |
|---|
| gamification |
| learning analytics |
| life course perspective |
| social informatics |
| social norms theory |
| theory of planned behavior |

**Figure 7:** Theories we expected to see in the literature

EmotionStore was a prototype also developed to address mental health and well-being Baras et al. (2016). In this prototype, students identified features that would be useful to facilitate taking care of one's mental health in a self-guided way. Some of these features are the ease of scheduling appointments with therapists, a minimal diary function using emoticons and emojis that can also be shared with friends on social media, an expanded diary for journaling, and tips organized into categories like sleep habits, working with teams, fatigue, and study habits during regular classes that can be updated based on the user's academic schedule. This app cannot be located at the time of writing.

Finally, a paper that was not yet available when we conducted our systematic review but is, nonetheless, important to include in this discussion is the ongoing development of CBCT by Williams et al. Williams et al. (2023). This application is in the prototype development phase. using evidence-based practices, the app adapts in-person cognitively-based compassion training (CBCT) to a mobile platform. This platform is also self-guided and will incorporate meditation, reminders, and easy engagement with peers and university counselors on the platform.

---

[6]https://www.moodprismapp.com/

These tools and interventions take similar approaches to address the needs of students in a self-guided way. The research objectives and outcomes will now be evaluated.

## 3.5. Research objectives and Outcomes

In examining the research objectives in the sample, several commonalities were identified.

All studies had at least one objective. 42.8% of studies presented a secondary objective, and 10.7% identified a tertiary objective. In these studies, three common themes for the objectives were identified. They are

· General Exploration

· Design and Develop

· Test and Improve

General Exploration is designed to gather the current state of the field. The most common themes in General Exploration are relationships, associations, impacts of factors like COVID-19, and the current state of students. Design and Develop consists of creating interventions, methods, and strategies. These include self-guidance models, best practices for faculty or student support, applications, and prediction models. The third theme Test and Improve, relates to methods as well as interventions. This includes measuring effectiveness, gamification, and user testing.

### 3.5.1. Implications

In addition to setting research in the context of the global COVID-19 pandemic, we see other implications for mental health issues emerge in studies about folks in the computing education pipeline. The findings of these studies occurred in two domains: university recommendations and application guidelines. Since the studies we examined for this review took place in a university setting, the findings indicated that the administration and faculty can do more to support students and address mental health needs. The largest emerging value was on co-creating a safe environment and identifying practices in the classroom with students Azmi et al. (2022), Siegel et al. (2022), Thiry and Hug (2021). In the literature, we see often-repeated challenges across disciplines. The omission on non-binary folks, for example, is a persistent issue because they are too few in number to yield any statistical significance. Studies that focus on just one marginalized group, like Hispanics Passos et al. (2022), are not necessarily reproducible across other groups. Social isolation is detrimental to success in computing classrooms, and there is still a lack of development in the design of intervention strategies,s elf-esteem, and self-efficacy. For application guidelines, Bakker and Rickard Bakker and Rickard (2019) point out that engagement is not necessarily an indicator for use.

### 3.5.2. Gender Based Implications

Most of the studies we reviewed did not find any specific results relating to gender. Sense of belonging, for example, and not belonging to a minority group pre-and post-COVID were not statistically significant. However, some gender implications are worth noting.

Bakker and Rickard identified a potential confounding influence of age and gender. Other gender-related findings include issues around mental health and anxiety. Reporting symptoms was generally higher for female students. Thiry and Hug found that women were disproportionately impacted as they were significantly more likely to experience multiple mental health challenges (X=36.238, df=12, p=.000) than men. Also commonly noted across studies were the presentation of higher anxiety, depression, and imposter syndrome scores in women than men. Also, female students with less social contact presented more depressive symptoms.

General observations were recorded that can be used to make departmental changes and create or reinformce initiatives. Gender imbalance and under-representation of other minority students in CS are issues the discipline faces, as noted by Grande. Passos et al. found that female CS students have higher mean number of factors that affected their well-being during the COVID-19 pandemic than male CS students. Finally, gender, personal history of sickness, and psychological health insurance offered by companies were identified as significant factors in determining mental health issues, and women often experienced more hate speech than men in online coding and social environments.

## 4. Discussion

In this section, we will present the salient features of this systematic literature review and we also provide certain interpretations of those features based on theoretical and practical knowledge from the field of adaptive learning, educational psychology, learning technologies and learning analytics. We divide the salient features into two groups. The first group represents the opportunities, while the second group represents the challenges in the field.

## 4.1. Opportunities

From the selected studies we distill four main opportunities in the field of mental well-being these opportunities include further analysis and new factors, interventions, longitudinal monitoring, wider and inclusive audience, replication of the studies with rich activities.

**Interventions:** There were a few suggestions based on the results and research outcome of the studies that prompted certain interventions to be conducted in the future work. For example , interventions based on self esteem and socialization and social networks could prompt healthy behaviors and attitudes Frieiro et al. (2022). Another set of prompts might be based on effective screening strategies Pelucio et al. (2022), virtual support systems Azmi et al. (2022), Soares Passos et al. (2020), self-care mental health systems and self-guided mental health care courses Penzenstadler et al. (2022). Some of the suggestions that focus specifically on the student populations, prompt the intervention to be based on the teaching methods and interactive classroom environments Soares Passos et al. (2020). Finally, mental health mobile apps could also be used to keep track of the current status and to be used in intervention whether and when appropriate Bakker et al. (2018), Bakker and Rickard (2018), Thach (2018),Kim et al. (2015). One of the key aspects in the potential success of these interventions is the objective evaluation and a consistent estimation of the construct that one is targeting.

**Further analysis and new factors:** The next opportunities stem from the papers are about new methods of analysing the data and considering the new factors to understand the processes underlying different levels of mental well-being. For example, there were suggestions about different types of screen activities on mental health in young adults Khouja et al. (2019) and understanding the impact of online hate speech on mental health Saha et al. (2019). Further suggestions included incorporating other factors for example personality traits, social interaction, mobility Baras et al. (2016), academic performance Kamarudin et al. (2022), and employee support Rosaline et al. (2022) into the analysis of factors that might have a relationship with the mental health of individuals and groups in computing. Further more there were suggestions related to the new methods of analysis in the field. For example understanding the causal relationship among concerned variables and measurements Rosenstein et al. (2020), using physiological measures and machine learning to conduct predictive analysis Rosaline et al. (2022), Kamarudin et al. (2022), and looking for mediators and moderators in some specific analysis. There should also be triangulations and mixed-methods analysis (involving interviews and focus groups) to have a better understanding of the aforementioned factors Mooney and Becker (2021), Grande et al. (2022).

**Longitudinal monitoring:** Another set of opportunities might have their roots in the longitudinal monitoring of the factors related to the mental health of computing professionals and students. For example, while studying the impact of recent pandemic on computing students' mental well-being, it is also important to monitor their well-being over time post pandemic Passos et al. (2022), Rao et al. (2021). Furthermore, we can also incorporate user sampling methods to assess social media usage and indicator of psychopathology data related to long term effects on mental well-being Faelens et al. (2019).

**Wider audience, replication, and inclusion:** the reported studies also point out at targeting wider audience for the interventions and standalone analyses Siegel et al. (2022), Grande et al. (2022), Soares Passos et al. (2020), Penzenstadler et al. (2022). This is proposed to gain a deeper understanding about various factors, that might be hidden, in a smaller and targeted population size. Another suggestion is also to perform replication studies with different demographic and geographical locations to gain a deeper understanding of the relationship between the mental health issues and various factors concerning computing professionals and students. Finally, the inclusion of richer activities and activities that are equitable and inclusive of a diverse range of participants to achieve higher levels of generalizability of the research outcome Thiry and Hug (2021), Mooney and Becker (2021), Yacoby et al. (2023). Specially, with the interventions and studies carried out in online settings, the engagement in the activities tend to decrease over time and therefore, it is necessary to design activities that are more engaging in long-term.

## 4.2. Challenges

**Low sample size:** most of the included contributions reported having lower sample size than what they targeted Mooney and Becker (2021), Kamarudin et al. (2022), Baras et al. (2016), Thach (2018). Not having enough samples, especially when conducting multi-institute studies Mooney and Becker (2021), might surface many confounding factors depending on the demographics and cultural differences among the institutes. Furthermore, the small sample size hinders not only in gaining a deeper understanding of phenomenon but also in generalization. Another limitation in the terms of sample size, especially when the studies are conducted within one institution is the low demographic diversity Kamarudin et al. (2022). This kind of shortcoming might hinder the generalization of the findings across a wider population group. Limited sample sizes also hinder in achieving the high accuracy and prediction capabilities when machine learning methods or least square regressions are applied Wang et al. (2019).

**Threats to construct validity:** On certain occasions, the construct validity was also not verified Soares Passos et al. (2020), Rosenstein et al. (2020), Saha et al. (2019), Wang et al. (2019). There could be multiple factors in affect for such issues for example, lack of qualitative data Rosenstein et al. (2020) to confirm the there was no construct overrepresentation (i.e., chosen measurement or operationalization of a construct is too narrow) or construct underrepresentation (i.e., when a measurement includes elements that are not part of the theoretical construct). If the proper descriptions of the categories are not provided Soares Passos et al. (2020), one risk hypothesis guessing (i.e., participants in an experiment can deduce the purpose of the study, they may alter their behavior intentionally). Moreover, if the proper statistical tool is not used Saha et al. (2019), then one risks the construct irrelevance variance (i.e., when sources of error variance or factors unrelated to the construct being measured affect the scores). Furthermore, with low sample size certain biases (e.g., sampling and/or cultural bias) can appear in the data impacting the construct validity. Finally, if the data is collected at very low sampling rate Yacoby et al. (2023) then it might also threaten the construct validity. Constructs can change over time, and if there is a significant delay between the measurement and the actual assessment of the construct, it can threaten validity.

**Limited statistical tools:** almost all the included contributions used the basic statistics and tools for their analysis. There were also only a few examples where mixed-methods were used (lack of diagnostic interviews can disrupt the meaningfulness of the outcomes Kim et al. (2015), Bakker et al. (2018), Bakker and Rickard (2018)). Sometimes the cross-sectional nature of the study does not permit causal conclusions Faelens et al. (2019). Also, there were no mediators or moderators considered in almost all the studies.

**Lack of generalizability:** there were several reasons mentioned in the included contributions that can impact the generalizability of the results and research outcomes. For example, if the contributions used only descriptive case studies Thiry and Hug (2021) then it is very difficult to reproduce and/or generalize. Similarly, if the sample sizes are small Mooney and Becker (2021), Kamarudin et al. (2022), Thiry and Hug (2021), Baras et al. (2016), Thach (2018) then the generalization is difficult. Moreover, if the demographic information is limited Grande et al. (2022), Thiry and Hug (2021) even then there might be an issue for generalization because in such cases, the demographics-based biases can not be reproduced. Finally, if the data is collected at very low sampling rate Yacoby et al. (2023) then it might not be feasible to replicate and generalized the interplay of the different constructs.

**Managing research in online setting**: some of the contributions included in this SLR had the studies conducted in online environments Siegel et al. (2022), Passos et al. (2022), Penzenstadler et al. (2022). This posed another set of challenges that might incur all aforementioned challenges. The participants might disengage and disrupt the data collection which might cause issues in the analysis. The participants might not understand the tasks clearly which poses challenges to construct validity. Moreover, the online participants might change their behaviour just because they know that they are being monitored, which might also impacts the construct validity. The behviour of participants might reflect their social desirability (i.e., participants may respond in ways that make them look more socially desirable, rather than providing honest responses), thus creating undesired biases in the data.

## 4.3. Limitations

Regarding limitations, the authors of this study had to make some methodological decisions (e.g., the choice of databases and the search query) that might introduce certain biases into the results that need to be accounted for. However, we did our best to avoid such biases by considering all the major databases as well as following the steps outlined by Kitchenham and Charters (2007). The second bias may be attributed to the selection of empirical studies and the coding of the papers in the papers themselves. It should be noted that the focus was clearly on the empirical evidence, and the coding of papers was done by three independent researchers who worked independently from each other. It has also been found that some elements of the papers were not accurately described, resulting in some missing information in the coding of the papers. The amount of missing information was minimal, and the results would not be affected significantly by the small amount of missing information.

# 5. Conclusions

From this exploration of the current state of research about mental health and well-being in computing education, the following findings were identified

- · Female computer science students report higher anxiety, depression, and imposter syndrome than men.

- · Technologies that are developed to address mental health in computing education are self-guided.

- · There is a lack of generalizability in case studies.

- · There is a lack of research on gender non-conforming students.

At the intersections of mental health and computing education, there is room for growth and development. Marginalized voices, if they do contribute to research, are often omitted from study results because they are too few in number. We, therefore, put forth a call to action to reflect critically on the research we do in mental health and the computing education pipeline so that our samples and designs are as diverse as the world we live in.